\documentclass[aps,prl,twocolumn, groupedaddress,superscriptaddress,amsfonts,amssymb,amsmath,citeautoscript,a4paper]{revtex4-1}
\pdfoutput=1

\usepackage[pdftex]{hyperref}
\hypersetup{colorlinks,
			linkcolor={blue!75!black!80!yellow},
			citecolor={blue!75!black!80!yellow},
			urlcolor={blue!75!black!80!yellow},
			pdfstartview=FitH}
\usepackage{graphicx}
\usepackage{xspace}
\usepackage{xr}
\usepackage{xcite}
\usepackage{xcolor}
\usepackage{stmaryrd} % provides \llbracket and \rrbracket for difference-operations
\usepackage[UKenglish]{babel}
\usepackage{placeins} % provides \FloatBarrier

\usepackage{siunitx}
\DeclareSIUnit[number-unit-product=]\percent{\char`\%} % remove space before percentage "units"

\usepackage{txfonts}  %Times Roman fonts
\usepackage{txfontsb} %Addition for txfonts, including old style numerals and greek

\makeatletter
\newcommand*{\addFileDependency}[1]{% argument=file name and extension
  \typeout{(#1)}% latexmk will find this if $recorder=0 (however, in that case, it will ignore #1 if it is a .aux or .pdf file etc and it exists! if it doesn't exist, it will appear in the list of dependents regardless)
  \@addtofilelist{#1}% if you want it to appear in \listfiles, not really necessary and latexmk doesn't use this
  \IfFileExists{#1}{}{\typeout{No file #1.}}% latexmk will find this message if #1 doesn't exist (yet)
}
\makeatother

\newcommand*{\myexternaldocument}[1]{%
    \externaldocument{#1}%
    \addFileDependency{#1.tex}%
    \addFileDependency{#1.aux}%
}

\makeatletter
\renewcommand\@make@capt@title[2]{%
	\@ifx@empty\float@link{\@firstofone}{\expandafter\href\expandafter{\float@link}}%
	\sffamily{\textbf{#1}}\@caption@fignum@sep#2
}%

\makeatother

\newcommand*\diff{\mathop{}\mathrm{d}}
\renewcommand{\Re}{\operatorname{Re}}
\renewcommand{\Im}{\operatorname{Im}}
\newcommand{\iu}{\mathrm{i}}
 
\newcommand{\x}{\textsubscript{\textit{x}}\xspace}
\newcommand{\para}{\parallel\xspace}
\newcommand{\rv}{\mathbf{r}}
\newcommand{\dpar}{d_{\parallel}}
\newcommand{\dperp}{d_{\perp}}
\newcommand{\jump}[1]{\llbracket #1 \rrbracket}
\newcommand{\nablav}{\boldsymbol{\nabla}}
\newcommand{\nv}{\hat{\mathbf{n}}}
\newcommand{\appropto}{\mathrel{\vcenter{
			\offinterlineskip\halign{\hfil$##$\cr
				\propto\cr\noalign{\kern1pt}\sim\cr\noalign{\kern-1pt}}}}}

\usepackage{textcomp} % for \textrightarrow
\usepackage{xifthen}
\usepackage{etoolbox}
\newboolean{togglecomments}
\newboolean{togglechanges} 

\setboolean{togglecomments}{true}
\setboolean{togglechanges}{false}

\newcommand{\comment}[2]{%
    \ifbool{togglecomments}%
    {\textcolor{blue!70!black}{\small\textsf{%
    \textsuperscript{\textsc{\textsf{\MakeLowercase{#1}}}}%
    [#2]}}} % if true, show comments
    {}}     % if false, do nothing
\newcommand{\swap}[2]{\ifbool{togglechanges}
    {#2}  % updates-only version
    {\textcolor{red!70!black}{[#1]}\textrightarrow{}\textcolor{green!50!black}{[#2]}}}
\newcommand{\remove}[1]{\ifbool{togglechanges}
    {}    % updates-only version
    {\textcolor{red!70!black}{#1}}}
\newcommand{\inset}[1]{\ifbool{togglechanges}
    {#1}  % updates-only version
    {\textcolor{green!50!black}{#1}}}
\newcommand{\optional}[1]{\ifbool{togglechanges}
    {}    % (minimal) updates-only version
    {\textcolor{yellow!50!orange!80!gray}{#1}}}

\newcommand{\citeremind}[1]{%
    [\textcolor{blue!75!black!80!yellow}{
        $\blacksquare$%
	    \ifthenelse{\isempty{#1}}
	        {}
	        {\textsuperscript{\tiny\textsf{#1}}}%
	}]\xspace}

\newcommand{\ie}{i.e.\@\xspace}  %Gobble-spaces of the "small" type ("small" via \@)

\newcommand{\eg}{e.g.\@\xspace}

\newcommand{\mitaffil}{\footnotesize Research Laboratory of Electronics, Massachusetts Institute of Technology, Cambridge, Massachusetts 02139, USA}
\newcommand{\bordeauxaffil}{\footnotesize Laboratoire Photonique Numerique et Nanosciences, Institut d'Optique d'Aquitaine, Universit\'{e} Bordeaux, CNRS, 33405 Talence, France}
\newcommand{\westlakeaffil}{\footnotesize 
Institute of Advanced Technology, Westlake Institute for Advanced Study, Hangzhou 310024, China}
\newcommand{\hunanaffil}{\footnotesize School of Physics and Electronics, State Key Laboratory of Advanced Design and Manufacturing for Vehicle Body, Hunan University, Changsha 410082, China}

\myexternaldocument{SI}

\begin{document}

%-----TITLE-----
\title{A General Theoretical and Experimental Framework for Nanoscale Electromagnetism}

%-----AUTHORS AND AFFILIATIONS-----
\author{Yi~Yang} 
\email{yiy@mit.edu}\email{dizhu@mit.edu}  \thanks{These authors contributed equally to this work.}
\affiliation{\mitaffil}
\author{Di~Zhu}	
\email{yiy@mit.edu}\email{dizhu@mit.edu}  \thanks{These authors contributed equally to this work.}
\affiliation{\mitaffil} 
\author{Wei~Yan} 				\affiliation{\bordeauxaffil} \affiliation{\westlakeaffil}
\author{Akshay~Agarwal}			\affiliation{\mitaffil} 
\author{Mengjie~Zheng} 			\affiliation{\mitaffil} \affiliation{\hunanaffil}
\author{John~D.~Joannopoulos} 	\affiliation{\mitaffil}
\author{Philippe~Lalanne}		\affiliation{\bordeauxaffil} 
\author{Thomas~Christensen}		\affiliation{\mitaffil}
\author{Karl~K.~Berggren} 		\affiliation{\mitaffil}
\author{Marin~Solja\v{c}i\'{c}} \affiliation{\mitaffil}

\begin{abstract}

    Local, bulk response functions, \eg permittivity, and the macroscopic Maxwell equations completely specify the classical electromagnetic problem, which features only wavelength $\lambda$ and geometric scales. 
    The above neglect of intrinsic electronic length scales $L_{\text{e}}$ leads to an eventual breakdown in the nanoscopic limit. 
    Here, we present a general theoretical and experimental framework for treating nanoscale electromagnetic phenomena.
    The framework features surface-response functions---known as the Feibelman $d$-parameters---which reintroduce the missing electronic length scales. 
    As a part of our framework, we establish an experimental procedure to measure these complex, dispersive surface response functions, enabled by quasi-normal-mode perturbation theory and observations of pronounced nonclassical effects---spectral shifts in excess of $\SI{30}{\percent}$ and the breakdown of Kreibig-like broadening---in a quintessential multiscale architecture: film-coupled nanoresonators, with feature-sizes comparable to both $L_{\text{e}}$ and $\lambda$.

\end{abstract}

\maketitle

\begin{figure*}[htbp]
	\centering
	\includegraphics[width=1\linewidth]{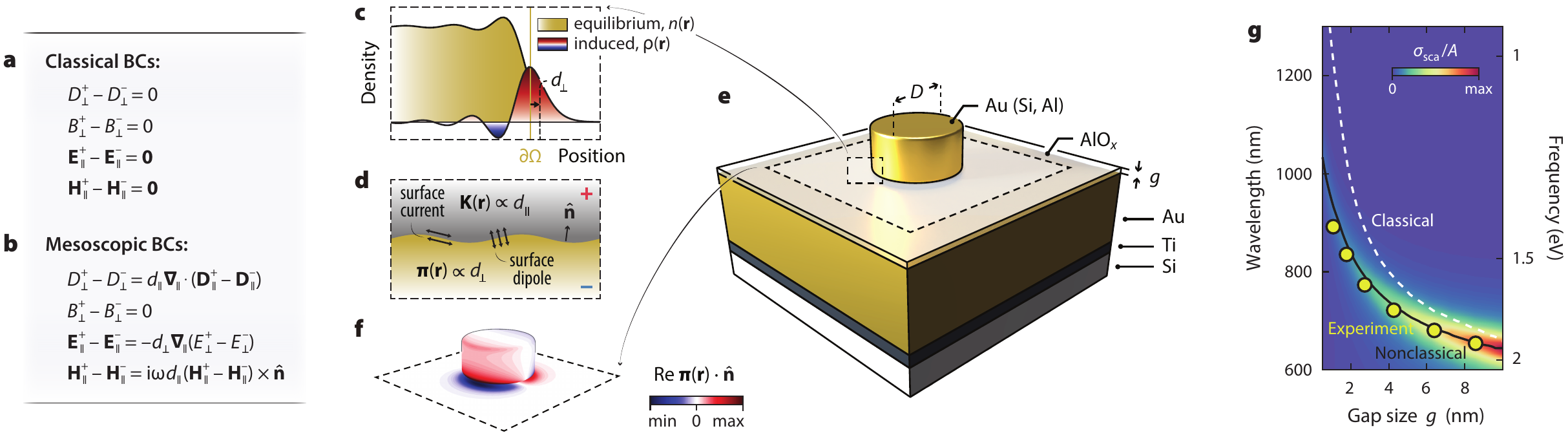}
	 \caption{%
	 	\textbf{Theoretical framework, experimental structure, and measured resonance frequencies versus theory.}
	 	\textbf{a.}~Classical and \textbf{b.}~mesocopic electromagnetic BCs.
	    \textbf{c.}~Equilibrium and induced densities, $n(\mathbf{r})$ and $\rho(\mathbf{r},\omega)$ (not to scale), at a jellium--vacuum interface (Wigner-Seitz radius, $r_s=\num{3.93}$; $\hbar\omega = \SI{1}{\eV}$) computed from (TD)DFT: $\dperp$ is the centroid of induced charge.
	 	\textbf{d.}~Nonclassical corrections can be formulated as self-consistent surface polarizations, representing effective surface dipole density $\boldsymbol{\pi}(\rv)$ and current density $\mathbf{K}(\rv)$. 
	 	\textbf{e.}~Schematic of the experimental structure: film-coupled Au nanodisks on an Au--Ti--Si substrate, separated by a nanoscale AlO\x gap $g$ (Si and Al nanodisks have also been studied).
	    \textbf{f.}~The nonclassical correction $\omega^{(1)}$ due to the $d$-parameters can be obtained from Eq.~\eqref{eq:qnm_correction}: the contribution from $d_\perp$ is proportional to the surface dipole density $\boldsymbol{\pi}(\rv)$, here shown for the $(1,1)$ gap plasmon of a film-coupled Au nanodisk ($D=\SI{63}{\nm}$, $g=\SI{4}{\nm}$).
	    \textbf{g.}~Observation of large nonclassical corrections (a spectral shift ${\gtrsim}\,\SI{400}{\nm}$) in film-coupled Au nanodisks ($D = \SI{63}{\nm}$).
	    Measured resonance frequencies of the $(1,1)$ plasmon blueshift (circles) relative to the classical prediction (dashed line) and quantitatively agree with our nonclassical calculations [solid line and intensity map (scattering efficiency $\sigma_{\text{sca}}/A$ where $A=\pi D^2/4$)].
	    }
	\label{fig1}
\end{figure*}

The macroscopic electromagnetic boundary conditions (BCs)---the continuity of the tangential $\mathbf{E}$- and $\mathbf{B}$-fields and the normal $\mathbf{D}$- and $\mathbf{H}$-fields (Fig.~\ref{fig1}a) across interfaces---have been well-established for over a century~\cite{maxwell1865viii}.
They have proven extremely successful at macroscopic length scales, across all branches of photonics. 
Even state-of-the-art nanoplasmonic studies~\cite{schuller2010plasmonics,halas2011plasmons,nielsen2017giant,benz2016single,chikkaraddy2016single,Akselrod:2014,kern2015electrically,Chen:2013,Henzie:2009}, exemplars of extremely interface-localized fields, rely on their validity.
This classical description, however, neglects the intrinsic electronic length scale associated with interfaces.
This omission leads to significant discrepancies between classical predictions and experimental observations in systems with deeply nanoscale feature-sizes, typically evident below ${\sim}\,\SIrange{10}{20}{\nm}$~\cite{Kreibig:1985, Cottancin:2006, Grammatikopoulos:2013, berciaud2005observation, Ciraci:2012, Charle:1998, Tiggesbaumker:1993, Scholl:2012, Raza:2013_Nanophotonics, iranzo2018probing}.
The onset has a mesoscopic character: it lies between the domains of granular microscopic (atomic-scale) and continuous macroscopic (wavelength-scale) frameworks.
This scale-delimited, mesoscopic borderland has been approached from above by phenomenological accounts of individual nonclassical effects---chiefly spill-out~\cite{Ozturk:2011, Zhu:2016, Skjolstrup:2018} and nonlocality~\cite{Boardman:1982a, David:2011, Raza:2015b, fernandez2012transformation, luo2013surface, Mortensen:2014, Khurgin:2015}---and from below, using explicit time-dependent density functional theory (TDDFT)~\cite{Zuloaga:2009a, Townsend:2011a, Stella:2013, Teperik:2013, Varas:2016}. 
The former approaches are uncontrollable and disregard quantities comparable to those they include; the latter is severely constrained by computational demands.
A practical, general, and unified framework remains absent.
Here, we introduce and experimentally demonstrate such a framework---amenable to both analytical and numerical calculations and applicable to multiscale problems---that reintroduces the missing electronic length scales.
Our framework should be generally applicable for modelling and understanding of any nanoscale (\ie all relevant length scales ${\gtrsim}\,\SI{1}{\nm}$) electromagnetic phenomena.

We reintroduce the electronic length scales by amending the classical BCs with a set of mesoscopic complex surface response functions, known as the Feibelman $\dperp$- and $\dpar$-parameters (Fig.~\ref{fig1}b)~\cite{Feibelman:1982, Liebsch:1997}: they play a role analogous to the local bulk permittivity, but for interfaces between two materials.
$\dperp$ and $\dpar$ are the missing electronic length scales---respectively equal to the frequency-dependent centroids of induced charge and normal derivative of tangential current at an equivalent planar interface (Fig.~\ref{fig1}c and Sec.~S1).
They enable a leading-order-accurate incorporation of nonlocality, spill-out, and surface-enabled Landau damping.

We start by summarizing the key elements of our framework: the $d$-parameters drive an effective nonclassical \emph{surface} polarization $\mathbf{P}_{\text{s}} \equiv \boldsymbol{\pi} + \iu\omega^{-1}\mathbf{K}$ (Fig.~\ref{fig1}d and Sec.~S2.A), with $\dperp$ contributing an out-of-plane surface dipole density $\boldsymbol{\pi} \equiv \dperp\varepsilon_0\jump{E_\perp}\nv$ and $\dpar$ an in-plane surface current density $\mathbf{K} \equiv \iu\omega\dpar\jump{\mathbf{D}_\para}$.
Here, $\jump{\mathbf{f}}\equiv \mathbf{f}^+ -  \mathbf{f}^-$ denotes the discontinuity of a field $\mathbf{f}$ across an interface $\partial\Omega$ with outward normal $\nv$; similarly,  $f_\perp\equiv \hat{\mathbf{n}}\cdot \mathbf{f}$ and $\mathbf{f}_\parallel \equiv (\mathbf{\hat{I}}-\mathbf{\hat{n}}\mathbf{\hat{n}}^{\mathrm{T}})\mathbf{f}$ denote the (scalar) perpendicular and (vectorial) parallel components of $\mathbf{f}$ relative to $\partial\Omega$.
These surface terms can be equivalently incorporated as a set of mesoscopic BCs for the conventional macroscopic Maxwell equations (also shown in Fig.~\ref{fig1}b and Sec.~S2.B):
\begin{subequations}\label{eqs:mesoscopic_bcs}
\begin{align}
    &\jump{D_\perp} 
    = -\iu\omega^{-1}\nablav_{\para}\cdot\mathbf{K}
    = \dpar\nablav_\para\cdot\jump{\mathbf{D}_\para},
    \label{eqs:mesoscopic_bcs_Dperp}
    \\
    &\jump{B_\perp} 
    = 0,
    \label{eqs:mesoscopic_bcs_Bperp}
    \\
    &\jump{\mathbf{E}_\para} 
    = -\varepsilon_0^{-1}\nablav_\para\pi = -\dperp\nablav_\para\jump{E_\perp},
    \label{eqs:mesoscopic_bcs_Epara}
    \\
    &\jump{\mathbf{H}_\para} 
    = \mathbf{K}\times \nv 
    = \iu\omega\dpar\jump{\mathbf{D}_\para}\times\nv.
    \label{eqs:mesoscopic_bcs_Hpara}
\end{align}%
\end{subequations}%
These mesoscopic BCs are a two-fold generalization from opposite directions.
First, they generalize the usual macroscopic electromagnetic BCs---$\jump{D_\perp} = \jump{B_\perp} = 0$ and $\jump{\mathbf{E}_\para} = \jump{\mathbf{H}_\para} = \mathbf{0}$---to which they reduce in the limit $\dperp=\dpar=0$. 
Second, they represent a conceptual and practical generalization of the Feibelman $d$-parameters' applicability---elevated from their original purview of planar~\cite{Feibelman:1982} and spherical~\cite{Apell:1982a} interfaces, and beyond recent quasistatic considerations~\cite{Christensen:2017}, to a fully general electrodynamic framework.

%--------------------------------------
%--------------------------------------

Experimentally, we establish a systematic approach to measure the $d$-parameter dispersion of a general two-material interface, and illustrate it using Au--AlO\x interfaces. 
While the $d$-parameters of simple metals can be accurately computed within jellium time-dependent density functional theory (TDDFT)~\cite{Feibelman:1982,Liebsch:1987}, $d$-parameters of noble metals, such as Au, require TDDFT beyond the jellium-approximation due to non-negligible screening from lower-lying orbitals~\cite{Liebsch:1993,Feibelman:1994,Christensen:2017}. 
We show that $d$-parameters can instead be measured experimentally: by developing and exploiting a quasi-normal-mode (QNM)-based~\cite{Lalanne:2018} perturbation expression, we translate these mesoscopic quantities directly into observables---spectral shifting and broadening---and measure them in designed plasmonic systems that exhibit pronounced nonclassical corrections.
Our experimental testbed enables a direct procedure to extract $d$-parameters from standard dark-field measurements, in a manner analogous to ellipsometric measurements of the local bulk permittivity.
Moreover, by investigating a complementary hybrid plasmonic setup, we discover and experimentally demonstrate design principles for structures that are classically robust---\ie exhibit minimal nonclassical corrections---even under nanoscopic conditions.

We briefly review the key nonclassical mechanisms that impact plasmonic response at nanoscopic scales~\cite{Feibelman:1982}.
First, equilibrium charge carriers spill out beyond the ionic interface~\cite{Lang:1970a}, blurring the classically-assumed step-wise transition between material properties; and second, dielectric response is nonlocal~\cite{PinesNozieres:1966, GiulianiVignale:2005}, \ie the $\mathbf{D}$- and $\mathbf{E}$-fields are related by a nonlocal response function $\varepsilon(\rv,\rv';\omega)$ rather than the local response function $\varepsilon(\rv,\omega)\delta(\rv-\rv')$ implicitly assumed in classical treatments 
(additionally, tunnelling~\cite{Esteban:2012, Savage:2012, Andersen:2013, Tan:2014, yan2015projected, Zhu:2016}  and size-quantization effects~\cite{Townsend:2011a, Townsend:2014, Halperin:1986}, ignored in this work, are non-negligible at feature-sizes below ${\approx}\,\SI{1}{\nm}$).
Individual aspects and consequences of these omissions have been studied extensively---\eg nonlocality~\cite{Boardman:1982a, David:2011, Raza:2015b, fernandez2012transformation, luo2013surface, Mortensen:2014, Khurgin:2015},
local-response spill-out~\cite{Ozturk:2011, Zhu:2016, Skjolstrup:2018}, and surface-enhanced Landau damping~\cite{Khurgin:2015}. 
However, to attain meaningful, quantitative comparisons of experiments and theory, a unified, general
framework that incorporates these mechanisms on equal footing is required.
In principle, TDDFT~\cite{Marques:2006} provides such a framework, but its range of applicability is limited to highly symmetric or sub-nanometric systems~\cite{Zuloaga:2009a,Townsend:2011a,Stella:2013,Teperik:2013,Varas:2016} due to prohibitive computational scaling.
Many promising electromagnetic systems, particularly plasmonic systems with multiscale features, are thus simultaneously incompletely described by macroscopic, classical electromagnetism and inaccessible to microscopic, quantum-mechanical frameworks like TDDFT.

\begin{figure}[tbp]
	\centering
	\includegraphics[width=1\linewidth]{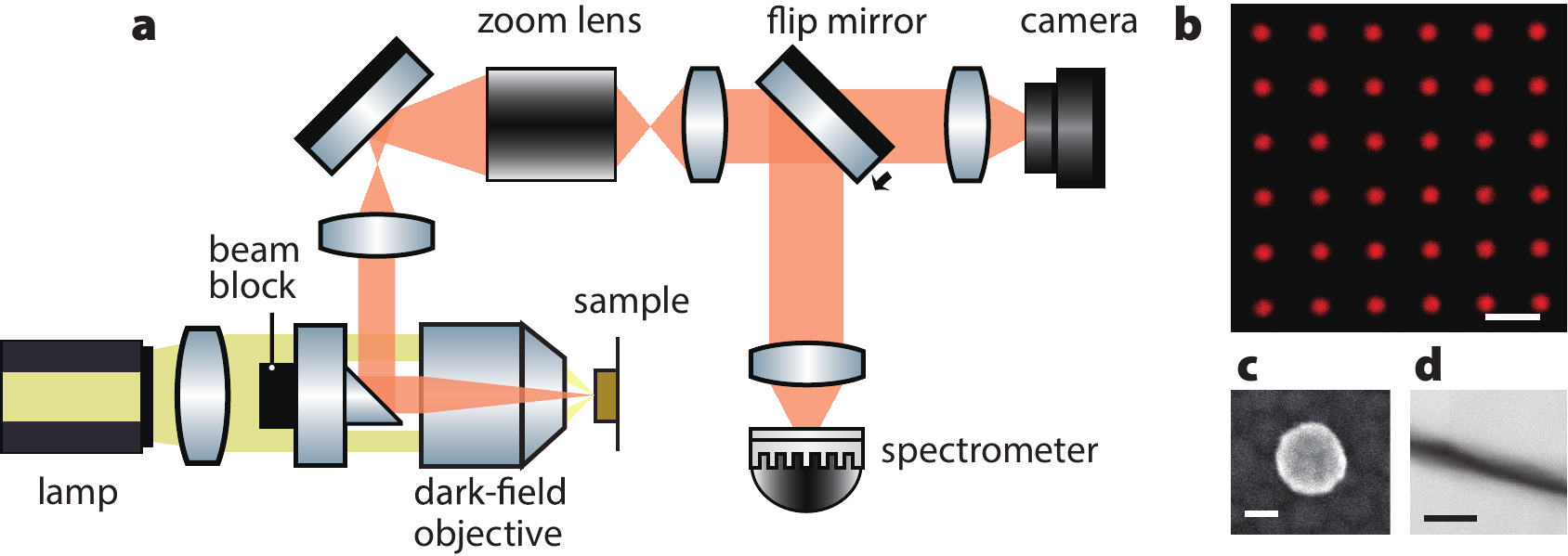}
	\caption{%
	 	\textbf{Schematic of measurement setup and micrographs of fabricated nanostructures.}
	 	\textbf{a.}~Tabletop dark-field scattering setup. It has a tunable magnification, and can record the dark-field image and measure the scattering spectrum (Sec.~S10).
	 	\textbf{b.}~Dark-field micrograph of a Au nanodisk array (scale bar, \SI{2}{\micro\meter}). 
	 	\textbf{c.}~SEM image of a single Au nanodisk (scale bar, \SI{40}{\nm}). 
	 	\textbf{d.}~Cross-sectional TEM image of an AlO\x gap (scale bar, \SI{10}{\nm}).} 
	\label{fig2}
\end{figure}

The extensive interest in film-coupled nanoresonators%
~\cite{moreau2012controlled,Akselrod:2014,chikkaraddy2016single,benz2016single, faggiani2015quenching, yang2017low}, which combine wavelength-scale resonators with a nanometric gap that approaches the intrinsic electronic length scale, is a particularly pertinent example that underscores the need for multiscale electrodynamic tools that incorporate nonclassical effects.
We designed and fabricated film-coupled nanodisks (Figs.~\ref{fig1}e and \ref{fig2}b--d) of various materials, to verify our framework and directly measure the $d$-parameters:
specifically, an optically-thick Au film (atop a Si substrate) is separated from lithographically defined Au, Si, or Al nanodisks (diameter, $D$) by a nanoscale AlO\x spacer, deposited by atomic layer deposition (ALD; see Sec.~S5), demarcating a film--nanodisk gap of thickness $g$.
Such film-coupled nanodisks support localized gap plasmon resonances~\cite{barnes2003surface}, which are $(m,n)$ integer-indexable according to their field variations in the azimuthal and radial directions, respectively~\citep{yang2017low}. 
The fundamental mode $(1,1)$ is optically accessible in the far field and exhibits highly confined electromagnetic fields within the gap, suggesting potentially large nonclassical corrections.

We implemented the mesoscopic BCs, Fig.~\ref{fig1}b, in a standard full-wave numerical solver~\citep{remark} (COMSOL Multiphysics; see Sec.~S3).
With specified $d$-parameters, this permits self-consistent calculations of \eg the nonclassical surface dipole density $\boldsymbol{\pi}(\rv)$, as shown in Fig.~\ref{fig1}f for the $(1,1)$ mode. 
Similarly, conventional electromagnetic quantities such as the scattering efficiency $\sigma_{\text{sca}}/A$ can be computed, enabling comparison with experiment (Fig.~\ref{fig1}g).
For Au disks, the $(1,1)$ resonance is consistently blueshifted relative to the classical prediction, with shifts exceeding \SI{30}{\percent} for the smallest considered gaps.

\begin{figure*}[tb]
	\centering
	\includegraphics[width=\linewidth]{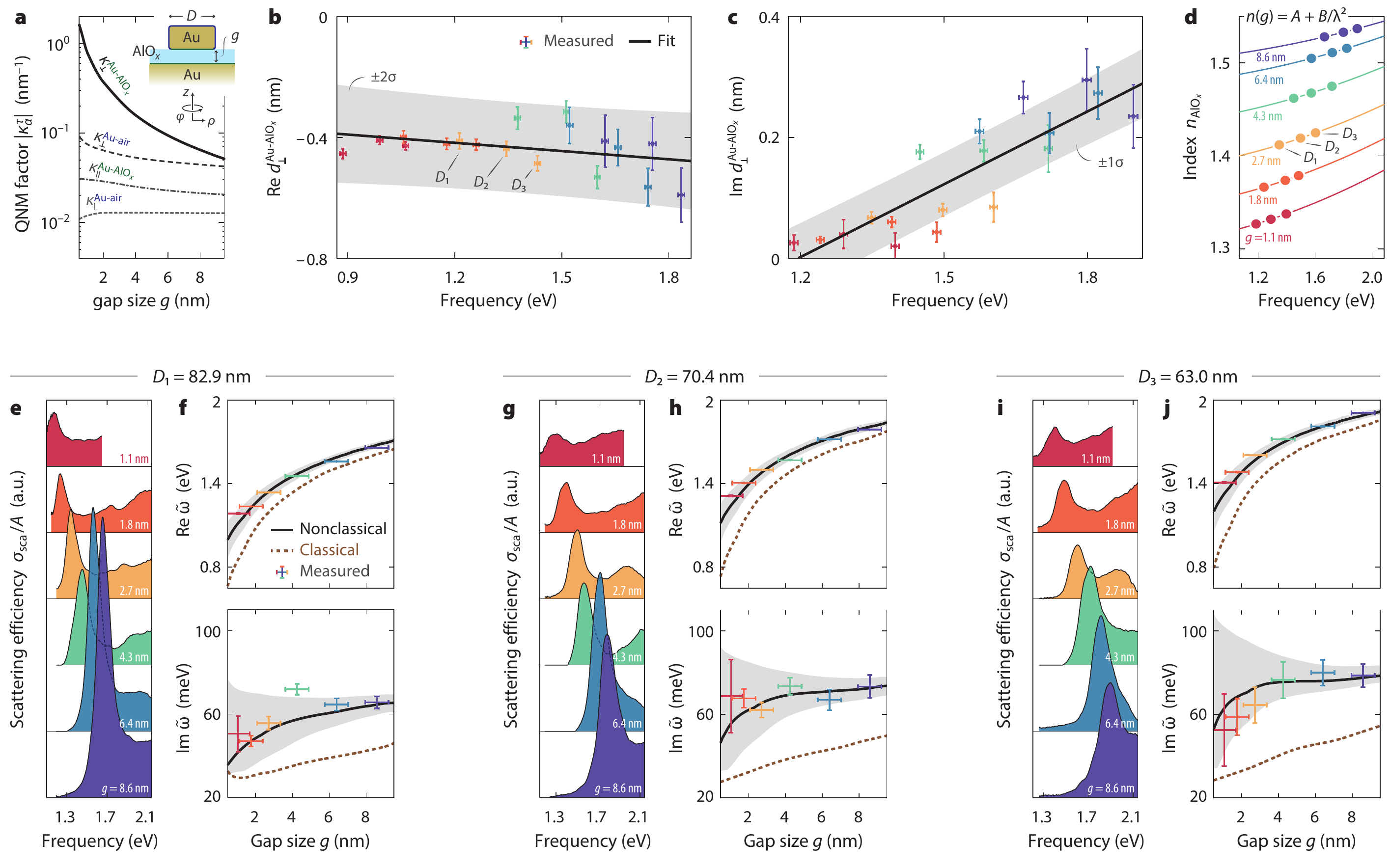}
	\caption{%
	 	\textbf{\boldmath Systematic measurement of the complex surface-response function $d_\perp(\omega)$ of the Au--AlO\x interface.}
	 	\textbf{a.}~Nonclassical perturbation strengths, calculated from QNM-based perturbation theory, Eq.~\eqref{eq:kappa_def}, in a film-coupled Au nanodisk (inset, $D= \SI{70.4}{\nm}$); 
	 	 $\kappa^{\text{Au--AlO\x}}$ is dominant. 
	 	\textbf{b--c.}~Measured (markers) dispersion of $\Re \dperp^{\text{Au--AlO\x}}$ (\textbf{b}) and $\Im \dperp^{\text{Au--AlO\x}}$ (\textbf{c}) and their linear fits (lines). Gap sizes are distinguished by color and diameters ($D_1 \approx \SI{82.9}{\nm}$, $D_2 \approx \SI{70.4}{\nm}$, and $D_3 \approx \SI{63.0}{\nm}$) decrement rightward.
	 	\textbf{d.}~ Measured thickness-dependent refractive indices of bare AlO\x films grown on Au.
	 	\mbox{\textbf{e--j.}}~Scattering efficiency (\textbf{e},\textbf{g},\textbf{i}) across distinct diameters and gap sizes and the extracted complex $(1,1)$ resonance eigenfrequencies (markers; \textbf{f},\textbf{h},\textbf{j}).
	 	While classical predictions (brown, dashed lines) deviate significantly from observations, our nonclassical calculations (black, solid lines;), employing the aforementioned linear $\dperp^{\text{Au--AlO\x}}(\omega)$ fit, are in quantitative agreement across all diameters.
	 	Shadings indicate fit-derived confidence intervals for our calculations; $2\sigma \approx \SI{95}{\percent}$ for $\Re \dperp^{\text{Au--AlO\x}}$ and $\Re\tilde{\omega}$ (\textbf{b},\textbf{f},\textbf{h},\textbf{j})
	 	and $1\sigma \approx \SI{68}{\percent}$ for $\Im \dperp^{\text{Au--AlO\x}}$ and $\Im\tilde{\omega}$ (\textbf{c},\textbf{f},\textbf{h},\textbf{j}).
	}
	\label{fig3}
\end{figure*}

To extract the surface response functions from observables, we develop a perturbation-theoretical description of the nonclassical spectral shift under the QNM framework~\cite{yang2015simple} with retardation explicitly incorporated:
the true eigenfrequency $\tilde{\omega} = \tilde{\omega}^{(0)} + \tilde{\omega}^{(1)} + \ldots$ (eigenindex implicit) exhibits a first-order nonclassical correction $\tilde{\omega}^{(1)}$ to its classical value $\tilde{\omega}^{(0)}$ (Sec.~S4)%
\begin{equation}
	\tilde{\omega}^{(1)}
	=
	\tilde{\omega}^{(0)}\sum_{\tau}\kappa_\perp^{\tau}d_\perp^{\tau} + \kappa_\parallel^{\tau}d_\parallel^{\tau},
	\label{eq:qnm_correction}%
\end{equation}%
with mode-, shape-, and scale-dependent nonclassical perturbation strengths (units of inverse length)
\begin{equation}\label{eq:kappa_def}
	\kappa_{\perp}^\tau
	\equiv 
	-\!\int_{\partial\Omega^\tau}\!\!\!
	\tilde{D}_{\perp}^{(0)} \llbracket\tilde{E}^{(0)}_{\perp}\rrbracket \diff^2\rv \\
	\text{\ \ and\ \,}
	\kappa_{\parallel}^\tau 
	\equiv
	\int_{\partial\Omega^\tau}\!\!\!
	\tilde{\mathbf{E}}_{\parallel}^{(0)}\cdot\llbracket \tilde{\mathbf{D}}^{(0)}_{\parallel} \rrbracket
	\diff^2\rv.
\end{equation}
Here, $\tau$ runs over all material interfaces such that $\bigcup_{\tau}\partial\Omega^{\tau} = \partial\Omega$, \ie $\tau\in\{\text{Au--AlO\x}, \text{Au--air}\}$ for our setup, while $\tilde{\mathbf{D}}^{(0)}$ and $\tilde{\mathbf{E}}^{(0)}$ denote the $\mathbf{D}$- and $\mathbf{E}$-fields of the (suitably normalized) classical QNM under consideration.
Figure~\ref{fig3}a shows the magnitudes of the nonclassical perturbation strengths in a film-coupled Au nanodisk: $\kappa_{\perp}^{\text{Au--AlO\x}}$, $\kappa_{\perp}^{\text{Au--air}}$, $\kappa_{\parallel}^{\text{Au--AlO\x}}$, and $\kappa_{\parallel}^{\text{Au--air}}$. 
Evidently, $\kappa_{\perp}^\tau$ far exceeds $\kappa_{\parallel}^\tau$ for all gap sizes of interest, rendering the impact of $\dpar^\tau$ negligible. 
Similarly, the impact of $\dperp^{\text{Au--air}}$ is negligible relative to $\dperp^{\text{Au--AlO\x}}$ since $\kappa_{\perp}^{\text{Au--AlO\x}}\gg \kappa_{\perp}^{\text{Au--air}}$ for small $g$; and more generally since $\dperp^{\text{Au--AlO\x}}>\dperp^{\text{Au--air}}$ due to screening from AlO\x (Sec.S16)~\cite{jin2015quantum}.
Jointly, this justifies the approximation
\begin{equation}
    \tilde{\omega}^{(1)}\simeq \tilde{\omega}^{(0)}\kappa_{\perp}^{\text{Au--AlO\x}}\dperp^{\text{Au--AlO\x}}.
    \label{eq:simpleshift}
\end{equation}
Inversion of Eq.~\eqref{eq:simpleshift} enables the direct experimental inference of $\dperp^{\text{Au--AlO\x}}$, given measured $\tilde{\omega}$ and calculated $\tilde{\omega}^{(0)}$ (since, to first order,  $\tilde{\omega}^{(1)}\simeq\tilde{\omega}-\tilde{\omega}^{(0)}$).
We note that $|\Re\kappa_{\perp,\parallel}^\tau|\gg|\Im\kappa_{\perp,\parallel}^\tau|$ (by \numrange[parse-numbers=false]{1}{2} orders of magnitude) for the considered gap-sizes: consequently, $\Re d_{\perp,\parallel}^\tau$ contributes to spectral shifting and $\Im d_{\perp,\parallel}^\tau$  to broadening.

%--------------------------------------
%--------------------------------------

We built a table-top dark-field microscope (Fig.~\ref{fig2}a), switchable between imaging and spectroscopy modes and with a 100\,--\,700$\times$ variable zoom, to measure $\tilde{\omega}$ from the optical response of the samples (Fig.~\ref{fig2}b).
Optical spectra were recorded at full zoom, capturing the scattered light from an ensemble of ${\lesssim}\, 100$ nanodisks (Sec.~S10).
Mutual coupling between nanodisks in the array is negligible, which is ensured by a lattice periodicity of \SI{2}{\micro\meter}, corresponding to an in-plane filling factor of less than \SI{1}{\percent}.
This allows an isolated-particle treatment.
The size distribution of the nanodisks was characterized systematically to adjust for the impact of inhomogeneous broadening in the measured scattering spectrum from the ensemble (see Fig.~\ref{fig2}c and Sec.~S7).
We measured the AlO\x gap size $g$ using a variable-angle UV-VIS ellipsometer and confirmed the results through cross-sectional transmission electron microscopy (TEM; see Fig.~\ref{fig2}d and Sec.~S8), finding good agreement with nominal ALD cycle expectations.
The Au substrate's surface roughness was measured to be $\approx\SI{0.6}{\nm}$ (RMS) using atomic force microscopy (AFM) and was taken as the gap size uncertainty.
Due to the conformal nature of the ALD~\citep{george2009atomic}, such roughness should have negligible influence on the scattering spectra, as we verified by numerical simulations (Sec.~S11).
These detailed characterizations eliminate the main sources of geometric uncertainty in the mapping between calculated $\tilde{\omega}^{(0)}$ and measured $\tilde{\omega}$, facilitating an accurate evaluation of the nonclassical shift $\tilde{\omega}-\tilde{\omega}^{(0)}$.

%--------------------------------------
%--------------------------------------

The scattering spectra of 18 Au nanodisk (height, $\SI{31}{\nm}$) arrays were collected (Figs.~\ref{fig3}e--j), spanning three diameters and six gaps sizes.
Associated complex eigenfrequencies $\tilde{\omega}$ were subsequently extracted by Lorenzian peak fitting (adjusting for inhomogeneous broadening by Voigt profile deconvolution; Sec.~S12). 
For the AlO\x spacer, we observed ellipsometrically---and include in our calculations---a thickness-dependent refractive index $n_{\text{AlO\x}}$ (Fig.~\ref{fig3}d and Sec.~S9), a commonly-observed effect in ultrathin ALD-grown AlO\x layers~\cite{groner2002electrical} and other ALD-grown materials~\cite{banerjee2018optical}.

Figures~\ref{fig3}bc show the complex surface-response function $\dperp^{\text{Au--AlO\x}}(\omega)$, extracted via Eq.~\eqref{eq:simpleshift}.
Within the considered spectral range, $\Re \dperp^{\text{Au--AlO\x}}$ (Fig.~\ref{fig3}b) reveal a nearly dispersionless surface response of comparatively large magnitude, from \SI{-0.5}{\nm} to \SI{-0.4}{\nm}. 
In contrast, $\Im \dperp^{\text{Au--AlO\x}}$ (Fig.~\ref{fig3}c) is strongly dispersive, increasing from ${\lesssim}\,\SI{0.1}{\nm}$ in the near-infrared to ${\approx}\,\SI{0.3}{\nm}$ in the visible.
The thickness dependence of $n_{\text{AlO\x}}$ imparts an attendant, implicit dependence to the inferred $\dperp^{\text{Au--AlO\x}}(\omega)$ (Sec.~S13). 
As a result, the frequency-fits in Figs.~\ref{fig3}bc convey a composite dependence along the $(\omega, n_{\text{AlO\x}})$-space (Fig.~\ref{fig3}d, circles) sampled by our data.

While the negative sign of $\Re\dperp^{\text{Au--AlO\x}}$---and the associated blueshift of $\Re(\tilde{\omega}-\tilde{\omega}^{(0)})$ (Figs.~\ref{fig3}fhj, top panel)---agrees with earlier observations in Au~\cite{Cottancin:2006, Grammatikopoulos:2013, berciaud2005observation} and Ag~\cite{Charle:1998, Tiggesbaumker:1993, Scholl:2012, Raza:2013_Nanophotonics} nanoparticles, the spectral shift is significantly larger.
There are two reasons: 
first, the nonclassical perturbation strength $\kappa_{\perp}^{\text{Au--AlO\x}}$ is much larger than in \eg standalone nanospheres or film-coupled nanospheres, due to strong field-confinement beneath the entire nanodisk footprint (Sec.~S4); 
and second, screening from the AlO\x cladding expels induced charge into Au, thereby enhancing $\dperp^{\text{Au--AlO\x}}$ relative to the unscreened interface, \ie relative to $\dperp^{\text{Au--air}}$ (Sec.~S16)~\cite{jin2015quantum}.

Nonclassical broadening due to surface-enhanced Landau damping, \ie $\Im (\tilde{\omega}-\tilde{\omega}^{(0)})$, is similarly enhanced for the same reasons (Figs.~\ref{fig3}fhj, lower panels). 
Classically, the linewidth reduces near-monotonically with gap-size, primarily due to increased light confinement (reduced radiative coupling). Instead, we observed---and predict, nonclassically---a near-constant broadening that is reduced slightly for very small gaps.
The near-constant broadening results from an interplay [Eq.~\eqref{eq:simpleshift}] among the strong (classical) gap-dependence of $\Re\tilde{\omega}$, the increase of nonclassical perturbation strength (Fig.~\ref{fig3}a) at smaller $g$, and the decrease of $\Im\dperp^{\text{Au--AlO\x}}$ towards the infrared (Fig.~\ref{fig3}c). 
Strikingly, the smallest gap does not produce the strongest nonclassical broadening (\ie $\Im\tilde{\omega}^{(1)}$), 
in contrast to the natural expectation of monotonically increasing $\Im\tilde{\omega}^{(1)}$ with decreasing $g$. 
Instead, $\Im\tilde{\omega}^{(1)}$ is minimal there---a consequence of the near-vanishing magnitude of the strongly dispersive $\Im\dperp^{\text{Au--AlO\x}}$ (Fig.~\ref{fig3}c). 
This behavior demonstrates the apparent breakdown of the empirical understanding of nonclassical broadening in nanostructures, known as Kreibig damping~\cite{Kreibig:1985}, which holds that $\Im\tilde{\omega}^{(1)} \appropto 1/g$.

%--------------------------------------
\begin{figure}[tbp]
	\centering
	\includegraphics[width=.95\linewidth]{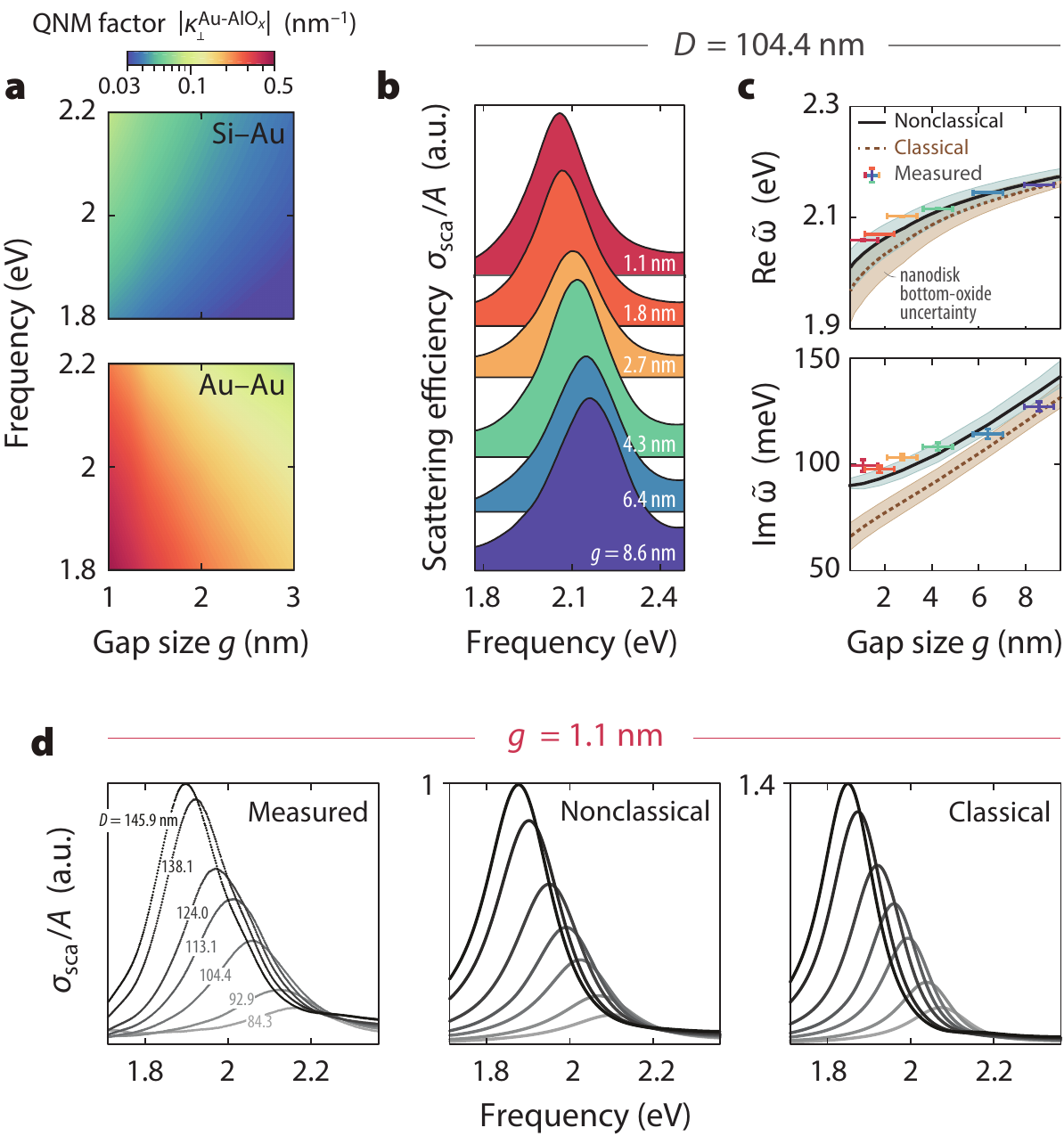}
	\caption{%
	 	\textbf{Robustness to nonclassical corrections.} 
	 	\textbf{a.}~The nonclassical perturbation strength is one order of magnitude smaller in the hybrid Si--Au system than in its Au--Au counterpart.
	 	 Si and Au nanodisk diameters are chosen to ensure spectral alignment of the $(1,1)$ resonance at every gap size (spanning $D\in[80,160]$\,\si{\nm} and $D\in[15,40]$\,\si{\nm}, respectively).
	 	\mbox{\textbf{b--d.}}~Observation of robust optical response in Si--Au setup with the detrimental quantum corrections mitigated. For a fixed diameter $D\approx\SI{104.4}{\nm}$, we obtain consistent, high-quality scattering spectra (\textbf{b}) showing only minor corrections in complex resonant frequencies (\textbf{c})---with identical outcomes observed across a wide range of nanodisk diameters, even at the smallest gap size (\textbf{d}).
	 	The nonclassical calculation for the Si--Au setup assumes $\dperp^{\text{Au--AlO\x}}=-\num{0.5}+\num{0.3}\text{i}\,\si{\nm}$, a constant extrapolation to higher frequencies from Fig.~\ref{fig3}b-c. 
	 	In \textbf{d}, measured and calculated spectra are normalized separately. Calculated spectra incorporate inhomogeneous broadening (${\approx}\,\SI{6}{\percent}$) due to disk-size inhomogeneity (Secs.~S7 and S12.B)
	 	}
	\label{fig4}
\end{figure}
%--------------------------------------

The observation of large nonclassical corrections in our coupled Au--Au setup frames a natural question: can nonclassical effects---which are often detrimental---be efficiently mitigated even in nanoscopic settings?
To answer by example, we consider a hybrid dielectric--metal design, replacing Au nanodisks with Si. 
Such hybrid configurations have been predicted to yield higher radiative efficiency with comparable overall plasmonic response~\cite{yang2017low} and have two key advantages for mitigating nonclassical effects:
first, undoped Si is effectively a purely classical material, \ie $d_{\perp,\parallel}^{\text{Si--AlO\x}}\simeq 0$, under the jellium approximation as it lacks free electrons; 
and second, high-index nanoresonators reduce field intensity at the metal interface while maintaining confinement in the gap region.
This hybridization can be exploited to reduce the nonclassical perturbation strength $\kappa_\perp^{\text{Au--AlO\x}}$ by an order of magnitude relative to that in the Au--Au design as shown in Fig.~\ref{fig4}a.
Our measurements confirm this prediction: for $D \approx \SI{104.4}{\nm}$ Si nanodisks, we observe a high-quality scattering spectra with a symmetric, single-resonance feature for all gap sizes (Fig.~\ref{fig4}b).
The measured resonance frequencies (Fig.~\ref{fig4}c) show only minor deviations from classical predictions, in both real and imaginary parts. 
While the inclusion of nonclassical effects improves the experimental agreement, the overall shift remains small, comparable to the uncertainties owing to the intrinsic oxide thickness beneath the Si nanodisk (Sec.~S12.C).
Considering a range of nanodisk diameters (Fig.~\ref{fig4}d), we reach an identical conclusion, even for the smallest considered gap (${\approx}\,\SI{1.1}{\nm}$): classical scattering spectra agree well with measurements, and nonclassical corrections are minor relative to those in the Au--Au system.
We found similar robustness across several additional gap sizes and diameters (Sec.~S14).

Equation~\eqref{eq:qnm_correction} suggests a complementary strategy for mitigating nonclassical effects:
if the sign of $\Re\dperp^\tau$ differs at distinct interfaces ($\tau$), the interface-summation $\Re\sum_\tau\kappa_\perp^\tau d_\perp^\tau$ will partially cancel.
While noble metals are known to spill outwards ($\Re\dperp>0$), simple metals, \eg Al, spill inwards ($\Re\dperp<0$)~\cite{Liebsch:1997}.
We found experimental evidence for such a partial cancellation in a combined noble--simple-metal setup (Al nanodisks above an Au substrate; Sec.~S15).

The mesoscopic framework presented here introduces a general approach for incorporating nonclassical effects in electromagnetic problems by a simple generalization of the associated BCs. 
Our experiments show how to directly measure the nonclassical surface-response functions---the Feibelman $d$-parameters---in general and technologically relevant plasmonic platforms.
Our findings establish the Feibelman $d$-parameters as first-class response functions of general utility. 
This calls for the compilation of databases of $d$-parameters at interfaces of photonic and plasmonic prominence, analogous and complimentary to the existing databases of local bulk permittivities~\cite{Palik:1997}.
In future work, our approach may also be extendable to cover two-dimensional materials.
Approaching the limits of plasmonic response---reached only at the nanoscale---inevitably requires an account of nonclassical effects; the tools developed here should enable that pursuit.

\FloatBarrier
% --- BACK-MATTER ---


\section*{Acknowledgements}
	We thank James Daley, Steven E.~Kooi, and Mark Mondol for assistance in sample fabrication and measurement.
	We thank Farnaz Niroui and Tony Zhu for generously lending us equipment.
	We thank fruitful discussions with Vladmir Bulovi\'{c}, Owen D.~Miller, and N.~Asger Mortensen. 
	We thank Paola Rebusco for critical reading and editing of the manuscript.
	This work was partly supported by the Army Research Office through the Institute for Soldier Nanotechnologies under contract No.\ W911NF-18-2-0048 and W911NF-13-D-0001, and Air Force Office of Scientific Research (AFOSR) grant under contract No.\ FA9550-18-1-0436.
	Y.\,{}Y. was partly supported by the MRSEC Program of the National Science Foundation under Grant No.\ DMR-1419807.
	D.\,{}Z. was supported by the National Science Scholarship from A*STAR, Singapore. 
	W.\,{}Y. and P.\,{}L. were supported by CNRS and Programme IdEx Bordeaux--LAPHIA (Grant No.\ ANR-10-IDEX-03-02).
	M.\,{}Z. was supported by the National Natural Science Foundation of China (Grant No.\ 11574078) the China Scholarship Council.
	T.\,{}C. was supported by the Danish Council for Independent Research (Grant No.\ DFF–6108-00667).

\section*{Author contributions}
	Y.\,{}Y. and T.\,{}C. conceived the idea.
	D.\,{}Z. fabricated the samples.
	Y.\,{}Y. and D.\,{}Z. designed the experiment, built the setup, conducted the scattering measurements, and performed the ellipsometry.
	T.\,{}C. derived the mesoscopic BCs.
	Y.\,{}Y., W.\,{}Y., and T.\,{}C. developed the numerical methods and Y.\,{}Y. performed the numerical calculations.
	W.\,{}Y. proposed the auxiliary potential method, performed density functional theory calculations, and implemented the QNM-based perturbation analysis.
	D.\,{}Z. performed the AFM measurement.
	A.\,{}A. and D.\,{}Z. performed the TEM measurement.
	D.\,{}Z. and M.\,{}Z. characterized nanoparticle size statistics.
	Y.\,{}Y., D.\,{}Z., W.\,{}Y., and T.\,{}C. analyzed the data.
	Y.\,{}Y., D.\,{}Z., and T.\,{}C. drafted the manuscript with extensive input from all authors.
	J.\,{}D.\,{}J., P.\,{}L., T.\,{}C., K.\,{}K.\,{}B., and M.\,{}S. supervised the project.

\section*{Competing interests}
	The authors declare no competing interests.


\begin{thebibliography}{65}
\bibitem [{rem()}]{remark}%
  \BibitemOpen
  \href@noop {} {}\bibinfo {note} {The COMSOL Multiphysics implementation will
  be made available upon acceptance of the manuscript.}\BibitemShut {Stop}%
\end{thebibliography}
\end{document}